\documentclass[prl,floatfix,twocolumn,amsmath,amssymb]{revtex4} 
\begin{document}   
\title{Transverse Observables and Mass Determination at Hadron Colliders} 
\date{September 18, 2007}
\author{Ben Gripaios}  
\email{ben.gripaios@epfl.ch} 
\affiliation{EPFL, BSP 720, 1015 Lausanne, Switzerland}
\affiliation{CERN, PH-TH, 1211 Geneva, Switzerland} 
\affiliation{Rudolf Peierls Centre for Theoretical Physics,
1 Keble Rd., Oxford OX1 3NP, UK}
\affiliation{Merton College, Oxford OX1 4JD, UK}
\begin{abstract}
I consider the two-body decay of a particle at a hadron collider into a visible and an invisible particle, generalizing $W \rightarrow e \nu$, 
where the masses of the decaying particle and the invisible decay particle are, {\em a priori}, unknown.
I prove that the transverse mass, when maximized over possible kinematic configurations, can be used to determine both
of the unknown masses. I argue that the proof can be generalized to cover cases such as decays of pair-produced superpartners
to the lightest, stable superpartner at the Large Hadron Collider.
\end{abstract}   
\maketitle 
Many particle physicists believe that there is new physics beyond the Standard Model, and that this new physics
will soon be probed by the Large Hadron Collider (LHC). New physics usually manifests itself in the presence
of new particles at higher energies. In order to understand the structure of the new physics, it is not enough
to simply discover the new particles; we must also measure their properties, such as mass, spin, and other quantum numbers.

The measurement of particle masses is typically very difficult in collisions of extended objects such as hadrons,
and indeed, discoveries of methods that enable us to do so have been of seminal importance in the history of particle physics.
As examples, I cite the Dalitz plot \cite{Dalitz:1953cp}, used to measure the masses of hadronic resonances, and the 
transverse mass observable, that allowed the first measurement of the $W$-boson mass \cite{Arnison:1983rp}.

The latter example is of particular interest here, because the relevant decay process involves an invisible
particle, the neutrino. In what follows, I will consider similar processes, in which the $W$ is replaced
by a new particle of unknown mass, and in which the invisible particle has unknown, but non-negligible mass.
Typical examples of such processes relevant for the LHC include decays of superpartners of 
known particles to the lightest, stable superpartner
(LSP), in the context of a supersymmetric completion of the Standard Model.
I present a method by which both of the unknown masses can be determined.

If this method could be shown to be experimentally viable, it would enable us to make measurements of the
absolute masses of superpartners at the LHC, including that of the LSP. As well as being of central importance to
particle physicists, such a measurement would be highly prized by the astrophysics community, for 
whom the LSP is the leading candidate 
for dark matter. If the LSP does make up the dark matter, then its mass, along with its relic density, play a 
fundamental r\^{o}le
in the large-scale evolution of the Universe.

Let me begin by recalling the example of the $W$-boson. Its mass can be measured at hadron colliders by maximizing the transverse mass observable.
To be concrete, consider a $W$ of mass $m$ and energy momentum $(E,p,q)$ in the laboratory frame, 
where $q$ is the momentum
in the beam direction and $p$ are components of the momentum in directions transverse to the beam. 
This $W$ decays into
a visible electron of mass $m_1$ and energy-momentum $(E_1,p_1,q_1)$ and an invisible neutrino of mass $m_2$ and energy-momentum 
$(E_2,p_2,q_2)$.
Simple kinematic considerations show that the transverse mass observable, defined by
\begin{gather} \label{f}
f = {m_1}^2 + {m_2}^2 - 2p_1 \cdot p_2 +2\sqrt{{p_1}^2 +{m_1}^2}\sqrt{{p_2}^2 +{m_2}^2},
\end{gather} 
is bounded above by the $W$ mass-squared, $m^2$. In what follows, it is convenient to define 
the transverse energy of the $W$ as $E'=\sqrt{p^2 +m^2}$,
and similarly for the electron and neutrino.
Now the neutrino is invisible, but its transverse momentum $p_2$ can be inferred with reasonable precision from the 
missing transverse
momentum in the detector. Furthermore, its mass $m_2$, though unknown, is negligible. 
Thus, for a given event, $f$ can be computed from data, and by maximizing
$f$ over a large sample of events, one can determine $m$ with good precision.

We shall shortly be entering the era of a new hadron collider, the LHC, and it is of interest to ask whether 
similar methods might enable us to measure the masses of the new particles we dearly hope to observe, for example
superpartners of the Standard Model particles \footnote{These are by no means 
the only particles whose masses we might hope to measure in this way, but I shall not discuss others here.}. 
In a realistic supersymmetric theory with conserved $R$-parity, for example, superpartners
typically decay into visible Standard Model states (like the electron above) and the lightest stable superpartner (LSP), 
which, like the neutrino, is invisible as far as the detector is concerned.

Unfortunately, we cannot simply carry over the method described above to measure the masses of superpartners decaying
to the LSP in this way, because the mass of the LSP is both unknown and non-negligible in general. 
There is a further complication,
coming from the fact that the conserved R-parity implies that superpartners are produced in pairs. So the decays of 
interest
involve two invisible LSPs, and, correspondingly, two unknown transverse momenta, whose sum is constrained to equal the missing
transverse momentum. This latter complication is not insurmountable: it turns out \cite{Lester:1999tx} that one can still define 
a suitable transverse mass variable, $M_{T2}$, that is bounded above by the mass of the decaying particle. By computing this observable
for a sample of events (taking into account all possible assignments of the unknown transverse momenta), one could still
measure the masses of decaying particles, if the mass of the LSP were known. 

The problem of our ignorance of the mass of the LSP remains.
Cho {\em et al.} \cite{Cho:2007qv} have recently given
evidence that this problem can be surmounted as well. They consider the special case of the decay 
of pair-produced gluinos
to quarks and a pair of LSPs. Their claim, which is based on a numerical simulation of events and an analysis of some
specific kinematic configurations, is that if $M_{T2}$ is considered as a function of the unknown mass $m_2$, then
it is continuous, but not differentiable (henceforth `it has a kink') exactly at the point where the mass equals the true mass \cite{Cheng:2007xv}.

This claim, if true, has remarkable implications: by identifying the kink on an experimental
plot, one would obtain measurements of the absolute masses of not one, but two superpartners, {\em viz.} the decaying
particle and the LSP. If experimentally viable, such a method would constitute a significant improvement in our ability to determine masses of new particles 
at the LHC, and to distinguish between candidate theories of physics beyond the Standard Model.

In this note, I should like to substantiate the claim of Cho {\em et al.}, by proving that a kink is present even in the simplest imaginable decay of this type, namely, the single-particle decay I discussed at the outset.
I shall claim that the generalization of the proof, which is based on high-school calculus, 
to the specific case considered by Cho {\em et al.}
and to other cases, should be straightforward. I will also show that, by measuring the gradient of the function in question
on either side of the kink, one can obtain an independent corroborative measurement of the two superpartner masses.

To make the proof as clear as possible, let me assume that there is just one, rather than two, 
direction transverse to the beam,
such that the tranverse momenta are one-vectors. 

I wish to maximise $f$ in (\ref{f}), but now considered as a function of some assumed mass $\tilde{m}_2$ 
for the invisible particle. 
Thus, I consider
\begin{gather} \label{f2}
f (\tilde{m}_2^2)= m_1^2 + \tilde{m}_2^2 - 2 p_1 p_2 +2\sqrt{(p_1^2 +m_1^2)(p_2^2 +\tilde{m}_2^2)},
\end{gather} 
maximised over all possible kinematic configurations. The possible energy-momenta are constrained by three energy-momentum
conservation conditions, {\em viz.}
\begin{align} \label{const1}
g_1 &\equiv E-E_1-E_2 = 0, \nonumber \\
g_2 &\equiv p-p_1 -p_2 =0, \nonumber \\
g_3 &\equiv q-q_1 -q_2 =0,
\end{align} 
together with three mass-shell conditions
\begin{align} \label{const2}
g_4 &\equiv E^2 - p^2 -q^2  - m^2 = 0, \nonumber \\
g_5 &\equiv E_1^2 - p_1^2-q_1^2 - m_1^2 = 0, \nonumber \\
g_6 &\equiv E_2^2 - p_2^2-q_2^2 - m_2^2 = 0.
\end{align}
Note that the mass-shell constraints involve the true mass, $m_2$, of the invisible decay particle.
To do the constrained maximization, I minimize $f+\lambda_i g_i$, subject to the constraints (\ref{const1}-\ref{const2}),
where the $\lambda_i$ are Lagrange multipliers.
I first note that the maximization with respect to $E_1$, $E_2$, $q_1$ and $q_2$ implies that
\begin{gather}
E_1 q_2 - E_2 q_1 = 0.
\end{gather} 
Now, the constraints (\ref{const1}) and (\ref{const2}), combined with this last equation yield the relation
\begin{gather} \label{const3}
m^2 = m_1^2 + m_2^2 - 2p_1 p_2 + 2 E'_1 E'_2
\end{gather}
at the maximum. 
(Note that the right-hand side is equal to $f ({\tilde{m}}_2^2=m_2^2)$, showing that $f$ is indeed maximised at $m^2$
when the assumed and true masses coincide, as I claimed earlier.)

Using (\ref{const3}), I can rewrite the expression for $f$ at the maximum, $\tilde{f}$, as
\begin{multline} \label{ill}
\tilde{f}({\tilde{m}}_2^2) =  m^2  + {\tilde{m}}_2^2 - m_2^2\\ + 2\sqrt{p_1^2 + m_1^2}\Big(\sqrt{p_2^2 + {\tilde{m}}_2^2} - \sqrt{p_2^2 + m_2^2}\Big),
\end{multline} 
where the $p_{1,2}$ are implicit functions of the various masses at the maximum.

I have thus far been rather cavalier in my treatment of $\tilde{f}$ and indeed, closer inspection of (\ref{ill})
shows that $\tilde{f}$ cannot obviously be regarded as a {bona fide} function of ${\tilde{m}}_2$ as it stands. 
The reason for this is that there are 
in fact many extrema of $\tilde{f}$ as defined in (\ref{f2}), and correspondingly many different possible values of $p_{1,2}$ in (\ref{ill}).
Though, as (\ref{const3}) shows, all these values lead to the same value for $\tilde{f}$, {\em viz.} $m^2$, when ${\tilde{m}}_2=m_2$, they do not lead to the 
same value for $\tilde{f}$ when ${\tilde{m}}_2 \neq m_2$. Thus, in maximizing $f$ away from the point ${\tilde{m}}_2=m_2$,
I must take care to choose the extremum that corresponds to the true maximum. In so doing, I obtain a single-valued
function.

Now I should like to argue that this prescription for constructing the function $\tilde{f}$ naturally
gives rise to an $\tilde{f}$ that has a kink at ${\tilde{m}}_2=m_2$. To see this, consider performing a Taylor expansion
of $\tilde{f}$, as written in (\ref{ill}), about the point ${\tilde{m}}_2=m_2$. One finds that
\begin{align}
\tilde{f}({\tilde{m}}_2^2) &= \tilde{f}(m_2^2)  + \tilde{f}'(m_2^2)({\tilde{m}}_2^2 - m_2^2) \nonumber \\
&= m^2 + ({\tilde{m}}_2^2 - m_2^2)\Big(1+\frac{E'_1}{E'_2}\Big).
\end{align} 
The first term is independent of which branch we choose, hence $\tilde{f}$ is $C^0$ at ${\tilde{m}}_2=m_2$; the second term
is not independent of which branch we choose, even though it is evaluated at ${\tilde{m}}_2=m_2$. To maximise $f$ in the neighbourhood
of ${\tilde{m}}_2=m_2$, we should choose the extremum
that gives the largest value of $1+\frac{E'_1}{E'_2}$ for ${\tilde{m}}_2 > m_2$, and we should choose the extremum that gives the smallest
value of $1+\frac{E'_1}{E'_2}$ for ${\tilde{m}}_2 < m_2$.

How do I find the extrema that give the largest and smallest values of $1+\frac{E'_1}{E'_2}$? Yet again, this is an
extremization problem. I wish to extremize $1+\frac{E'_1}{E'_2}$, subject to the constraint (\ref{const3}).
To do so in the most efficient manner, I first write the constraint (\ref{const3}) purely in terms of the transverse energies, $E'_{1,2}$, as
\begin{gather} \label{const4}
{E'}_1^2 m_2^2 +{E'}_2^2 m_1^2 - 2M^2 E'_1 E'_2 +M^4 -m_1^2 m_2^2 = 0.
\end{gather} 
Here I have defined $2M^2 = m^2 - m_1^2 -m_2^2$ for convenience. Note that $M^2$ is positive semi-definite above the mass threshold for the decay.

Now extremize $1+\frac{E'_1}{E'_2}$, including the constraint (\ref{const4}) via a Lagrange multiplier.
Upon eliminating the Lagrange multiplier, one obtains the condition that, for an extremum,
\begin{gather}
{E'}_1^2 m_2^2 + {E'}_2^2 m_1^2 - 2M^2E'_1E'_2 = 0.
\end{gather} 
From this it is straightforward to obtain the extremal values of $1+\frac{E'_1}{E'_2}$; they are
\begin{gather} \label{grads}
1+\frac{E'_1}{E'_2} = 1 + \frac{M^2 \pm \sqrt{M^4 -m_1^2 m_2^2}}{m_2^2}.
\end{gather} 
Comparing with (\ref{const4}), we see that these extrema are in fact obtained asymptotically, at large energies.

These results are all easy to understand. The constraint (\ref{const4}) is just the constraint that one would obtain if 
one considered
a two-body decay process in 1+1 spacetime dimensions, with $E'_{1,2}$ corresponding to the 
true, rather than transverse, energies.
The kinematics of such a decay is completely fixed in the rest frame of the decaying particle, and the only freedom
in the problem comes from the freedom to boost the decaying particle's rest frame with respect to the laboratory frame. 
The maximum in $1+\frac{E'_1}{E'_2}$ is obtained asymptotically by making an arbitrarily large boost in the direction of
$p_1$ in the lab frame, and the minimum is obtained by making a boost in the opposite direction, namely that of $p_2$. 
It is, moreover, clear that the situation is symmetric with respect to 
particles 1 and 2 and this symmetry is manifest in the extrema: the maximal value of $1+\frac{E'_1}{E'_2}$ is the same as the 
minimal value of $1+\frac{E'_2}{E'_1}$ and {\em vice versa}.

The two extremal values of $1+\frac{E'_1}{E'_2}$ do not coincide unless one sits exactly on the mass threshold, that
is for $m=m_1 + m_2$. Thus we have proven that the function $f$, when maximised over the possible kinematic configurations, is continuous,
but not differentiable, for all values of the masses above, but not at, the threshold for the decay. 
Thus, the absolute masses $m$ and $m_2$ can be determined in experiment simply by maximising $f$, 
considered as a function of the assumed mass ${\tilde{m}}_2$, over a suitably large number of events.
The function should contain a point that is continuous, but not differentiable, and the co-ordinates of this point 
are $(m_2^2,m^2)$ \footnote{I have only shown the existence of one such point. It is not impossible that other such points may exist for the function $f$.}.
The difference in the gradients on either side of the point increases as one moves 
further away from the decay threshold. It would appear, therefore, that the special point would be most easily identified
experimentally in cases where the decay is well above threshold. 

I remark that measurement of the two gradients of the function $\tilde{f}$, which are given by (\ref{grads})
would enable an independent determination of the two masses to be performed. I do not know whether it will be possible to measure these
accurately in practice.

It is, perhaps, amusing to add that this method, applied to the case of $W \rightarrow e \nu$, would enable a laboratory
measurement of the absolute mass of neutrino, or at least an upper bound thereon. I suspect, however, that the measurement
would not be a very precise one.

Lastly, let me argue that the generalization of the proof given here to the case considered by Cho {\em et al.} and other cases
is not too difficult. It is clear that the key element of the proof is the assertion that the gradients
of the function $\tilde{f}$ do not match on either side of true mass point. The reason this occurs is simply because
$f$ has extrema which are degenerate at the true mass point, but not elsewhere. Exactly the same phenomenon occurs in the more complicated
cases; the only difficulty is that the possible energy-momentum configurations over which one must maximize are more involved. 
A forthcoming publication \cite{BGL} will supply a much more general proof, as well as Monte Carlo simulations suggesting that the method is feasible at the LHC.
\begin{acknowledgments}
I thank A.\ Barr, J.\ J.\ Binney, R.\ Rattazzi and S.\ M.\ West for useful discussions.
\end{acknowledgments}

\end{document}